\documentclass[prl,twocolumn,aps,superscriptaddress]{revtex4-1}
\usepackage{graphicx,graphics,color,epsfig}% Include figure files
\usepackage{times}
\usepackage{epstopdf}
\usepackage{bm}
\usepackage{amsmath}
\usepackage{amssymb}
\usepackage{diagbox}
\usepackage{color}
\usepackage{appendix}
\usepackage{dcolumn}% Align table columns on decimal point
\usepackage[colorlinks=true, letterpaper=true, pdfstartview=FitV, linkcolor=blue, citecolor=blue, urlcolor=blue]{hyperref}

\begin{document}

\title{Electrically Controllable Chiral Phonons in Ferroelectric Materials}

\author{Hao Chen}
\affiliation{Department of Physics, University of Science and Technology of China, Hefei 230026, China}
\affiliation{Phonon Engineering Research Center of Jiangsu Province, Center for Quantum Transport and Thermal Energy Science, Institute of Physics Frontiers and Interdisciplinary Sciences, School of Physics and Technology, Nanjing Normal University, Nanjing 210023, China}

\author{Weikang Wu}
\affiliation{Key Laboratory for Liquid-Solid Structural Evolution and Processing of Materials,
	Ministry of Education, Shandong University, Jinan 250061, China}
\affiliation{Research Laboratory for Quantum Materials, Singapore University of Technology and Design, Singapore 487372, Singapore}

\author{Kangtai Sun}
\affiliation{Department of Physics, National University of Singapore, Singapore 117551, Republic of Singapore}

\author{Shengyuan A. Yang}
\email{shengyuan\_yang@sutd.edu.sg}
\affiliation{Research Laboratory for Quantum Materials, Singapore University of Technology and Design, Singapore 487372, Singapore}

\author{Lifa Zhang}
\email{phyzlf@njnu.edu.cn}
\affiliation{Phonon Engineering Research Center of Jiangsu Province, Center for Quantum Transport and Thermal Energy Science, Institute of Physics Frontiers and Interdisciplinary Sciences, School of Physics and Technology, Nanjing Normal University, Nanjing 210023, China}

\begin{abstract}
Chiral phonons have attracted increasing attention, as they play important roles in many different systems and processes. However, a method to control phonon chirality by external fields is still lacking. Here, we propose that in displacement-type ferroelectric materials, an external electric field can reverse the chirality of chiral phonons via ferroelectric switching. Using first-principles calculations, we demonstrate this point in the well known two-dimensional ferroelectric In$_2$Se$_3$. This reversal may lead to a number of electrically switchable phenomena, such as chiral phonon induced magnetization, phonon Hall effect, and possible interface phonon modes at ferroelectric domain boundaries. Our work offers a new way to control chiral phonons, which could be useful for the design of thermal or information devices based on them.
\end{abstract}

\maketitle

\textcolor{blue}{\textit{Introduction} ---}
Phonons are quantized vibration modes of a crystal lattice. A recent development in the study of phonons is the recognition that certain phonon modes may have a definite chirality~\cite{zhang2015chiral,zhu2018observation,ueda2023chiral}, i.e., they differs under parity operation. Such chiral phonons can carry nonzero phonon angular momentum and magnetic moment. Along rotation axis, chiral phonons may also have quantized pseudo-angular momentum (PAM), which manifests in selection rules of optical transitions~\cite{li2019momentum,he2020valley,liu2020multipath,liu2021nonlinear}. Recent studies also revealed propagating chiral phonons in bulk crystals~\cite{chen2021propagating} and the coupling between chiralities of phonons and lattice structure in a chiral crystal~\cite{chen2022chiral}. So far, chiral phonons have been predicted and studied in many material systems~\cite{bistoni2021intrinsic,suri2021chiral,li2021topological,sonntag2021electrical,zhang2022chiral,wang2022chiral}. The concept also plays important roles in diverse fields, such as function of biomolecules~\cite{choi2022chiral}, design of novel phononic devices~\cite{chen2022chiral}, activating spin Seebeck effect~\cite{kim2023chiral}, phonon-magnon hybridization~\cite{cui2023chirality}, ultrafast demagnetization~\cite{tauchert2022polarized}, and etc.

To realize the full potential of chiral phonons for future applications, it is necessary to have a way to manipulate their chirality. Especially, for device applications, it is preferred that such manipulations are achieved by applied electric or magnetic fields. The magnetic field breaks time reversal symmetry and hence potentially can distinguish different chiralities. However, to achieve this, the required magnetic field must be huge, due to the generally weak coupling between magnetic field and phonons. Then, how about electric field? Since phonons are tied with the crystal structure, a natural thought is to request the $E$ field to induce a change in the structure. Fortunately, the displacement-type ferroelectric materials can meet this condition. Using $E$ field, their structures can be switched between configurations with different electric polarizations, due to displacement of atoms in the structure. This opens an opportunity to control phonons via electric means.

In this work, we explore this idea of electric control of chiral phonons using ferroelectric crystals.
By first-principles calculations, we successfully demonstrate the idea in a well known two-dimensional (2D) ferroelectric
In$_2$Se$_3$. We show that the ferroelectric switching by an applied $E$ field simultaneously flip the chirality of phonons.
The quantized PAM of chiral phonons at $K$ and $K'$ points and the phonon Berry curvature are also reversed in the process.
This effect leads to several electrically controllable chiral phonon phenomena, including thermally induced phonon magnetization, phonon Hall effect, and possible interface phonon mode at ferroelectric domain boundaries. Our discovery opens a route to control the chirality of phonons, which may serve as basis for designing novel devices based on chiral phonons.

\textcolor{blue}{\textit{General Idea} ---}
We start with a simple analysis of the requirements on the crystal system to achieve our idea.  First of all, time reversal ($\mathcal{T}$) and inversion ($\mathcal{I}$) operations have opposite effects on phonon chirality. Hence, to have a net phonon chirality, one of the two symmetries must be broken in the host crystal. As mentioned, the effect of $\mathcal{T}$ breaking (such as applied $B$ field or magnetic ordering) on the phonon sector is generally rather weak, so the research is focused on systems with broken $\mathcal{I}$. In this regard, ferroelectrics are actually a very natural choice, since a fundamental character of them is also $\mathcal{I}$ symmetry breaking.

Second, another defining property of ferroelectrics is that their polarization states can be switched by applied $E$ field. For displacement type ferroelectrics, the two states correspond to two crystal structures connected by $\mathcal{I}$ operation. Under $\mathcal{I}$, the phonon mode $\psi_{n\bm q}$ for a given momentum $\bm q$ and branch index $n$ must get a reversed chirality (circular polarization). Thus, in ferroelectrics, there is a strong coupling between electric polarization and phonon chirality, and ferroelectrics offer a good platform for manipulating chiral phonons.

Third, if we want to have chiral phonons with quantized PAM, we may impose an additional requirement that the ferroelectric has an $n$-fold axis with $n\geq3$. Then, in momentum space, chiral phonons located at the $C_n$ invariant high-symmetry points/paths will acquire quantized PAM. Again, under ferroelectric switching,
PAM for these phonon modes must also be flipped.

\begin{figure}[htbp]
\includegraphics[width=1\linewidth]{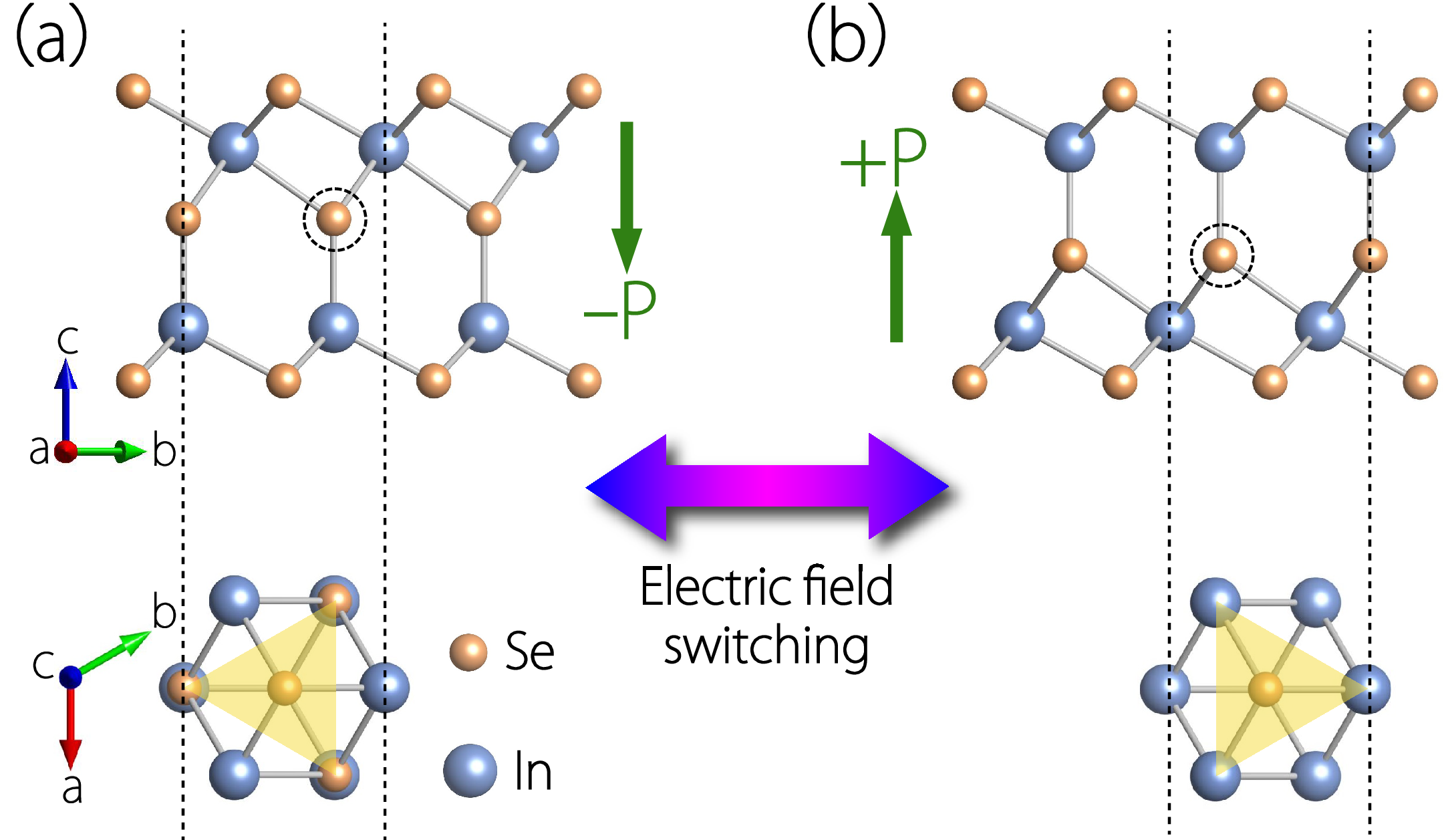}
\caption{\label{fig1} Lattice structure of 2D In$_2$Se$_3$. (a), (b) Side and top view of two energy-degenerate In$_2$Se$_3$ structures with opposite out-plane spontaneous electric polarization $P$, and the external electric field can switch the polarization $P$. The dotted circles indicate the Se atoms in the middle layer, and the flipping of the ferroelectric polarization $P$ is accompanied by the displacement of this atom. The definitions of $-P$ and $+P$ states are adopted from ref.~\cite{ding2017prediction}}.
\end{figure}

\textcolor{blue}{\textit{Electrically Controllable Chiral Phonons in Ferroelectric In$_2$Se$_3$} ---}
With the above considerations, we choose to demonstrate our proposal in a well known example of 2D ferroelectrics: monolayer In$_2$Se$_3$. In$_2$Se$_3$ is a van der Waals layered material in the bulk form. Its 2D ferroelectricity in the monolayer limit was predicted by Ding et al. in 2017~\cite{ding2017prediction} and was successfully verified in experiment~\cite{cui2018intercorrelated,xue2018rooom}. Because of this interesting property, 2D In$_2$Se$_3$ has attracted great research interest in the past few years.

Figure.~\ref{fig1} shows the crystal structure of monolayer ($\alpha$-phase) In$_2$Se$_3$. It consists of five atomic layers, as arranged in the sequence of Se-In-Se-In-Se. A primitive unit cell contains two In atoms and three Se atoms, as marked in Fig~\ref{fig1}. The crystal has a space group of $P3m1$ (No.~156), without $\mathcal{I}$ symmetry. It is noted that the structure actually respects $\mathcal{I}$ if the middle-layer Se atoms are removed. Hence, the $\mathcal{I}$ symmetry here is due to the arrangement of the middle-layer Se atoms. There are two energy degenerate configurations, as shown in Fig.~\ref{fig1}. The whole middle Se layer is shifted either up (Fig.~\ref{fig1} (a)) or down (Fig.~\ref{fig1} (b)). This results in an out-of-plane electric polarization $P$, and the two configurations have opposite $P$, connected by $\mathcal{I}$ operation. The ferroelectric switching just corresponds to the displacement of the middle-layer Se atoms.
Experimentally, the two polarization states can be switched by an applied vertical $E$ field on the order of 1 V/nm~\cite{cui2018intercorrelated,xue2018rooom}.

\begin{figure*}[htbp]
\includegraphics[width=7.0 in,angle=0]{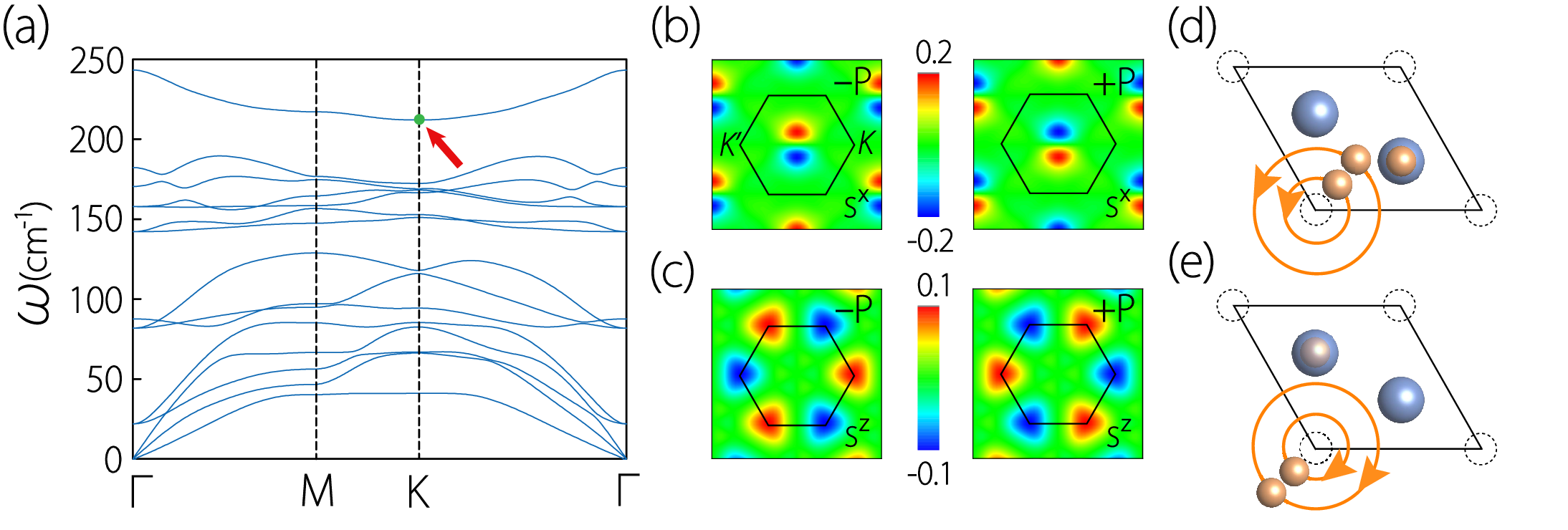}
\caption{\label{fig2} (a) In$_2$Se$_3$ with opposite electric polarization $P$ has the same phonon spectra. (b) shows the $x$-direction phonon polarization of the top (fifteenth) branch for the $-P$ and $+P$ In$_2$Se$_3$, respectively. (c) shows the $z$-direction phonon polarization of the top (fifteenth) branch for the $-P$ and $+P$ In$_2$Se$_3$, respectively. The high-symmetry points $K$ and $K'$ are marked. (d) and (e) show the oscillation pattern (top view) for the mode $\psi_{15,K}$ for the $-P$ and $+P$ In$_2$Se$_3$, respectively. The red arrow in (a) indicate this mode.}
\end{figure*}

Now we turn to the phonon spectrum of 2D In$_2$Se$_3$. Figure.~\ref{fig2} (a) shows the result obtained from first-principles calculations. It is important to note that the two polarization states give the identical dispersions, as it should be. There are three acoustic branches and twelve optical branches, corresponding to the five atom basis in a unit cell. The maximal phonon frequency goes up to over 200 cm$^{-1}$. Although the two polarization states have the same phonon spectra, as we show below, the phonon modes (wave functions) in fact differ. This difference is manifested in important characters, such as phonon chirality, PAM, and phonon Berry curvature.

First, let's investigate the phonon chirality in 2D ferroelectric In$_2$Se$_3$. The chirality of a phonon mode is defined in terms of its circular polarization. Considering the five atoms in the primitive cell, the phonon mode eigenvector can be expressed as a $3\alpha$ $(\alpha=5)$ dimensional vector $\psi$, describing the oscillation of atoms. The phonon circular polarization operator is defined as $\hat{S}^i\equiv\sum_{\alpha=1}^{5}(|R^i_{\alpha}\rangle\langle R_{\alpha}^i| - |L_{\alpha}^i\rangle\langle L_{\alpha}^i|)$, where $i=x,y,z$ label the three spatial directions, $\alpha$ label the atoms in a unit cell, and $|L^i_\alpha\rangle$ ($|R^i_\alpha\rangle$) is the left (right) handed circular polarization basis in the $i$ direction at atomic site $\alpha$.
Then, the circular polarization for the phonon mode $\psi_{n\bm q}$ can be obtained as
\begin{eqnarray}\label{eq:polarization}
S^{i}_{n\bm q} = \hbar\langle\psi_{n\bm q}|\hat{S}^{i}|\psi_{n\bm q}\rangle.
%= \hbar\sum_{\alpha=1}^{5}(|\epsilon_{R_{\hat{n},\alpha}}|^{2} - |\epsilon_{L_{\hat{n},\alpha}}|^{2})
\end{eqnarray}
Here, $\hbar$ is added such that $S^{i}_{n\bm q}$ directly gives the $i$-component of angular momentum of this phonon mode, and $S^{i}_{n\bm q}<0$ ($>0$) tells us the phonon mode $\psi_{n\bm q}$ is left (right) handed along the $i$-th direction.
Figure.~\ref{fig2} (b) and (c) show the calculated phonon circular polarization for the top branch ($n=15$). One observes that the $S^i$ values flip sign for the two polarization states (labeled as $-P$ and $+P$ as in Fig.~\ref{fig1}), indicating the reversal of phonon chirality under ferroelectric switching. This chirality control can also be visually seen in Fig.~\ref{fig2} (d) and (e), where we schematically plot the oscillation pattern (top view) for the mode $\psi_{15,K}$, i.e., the mode indicated by the red arrow in Fig.~\ref{fig2} (a). The chirality clearly flips between the two polarization states. These results confirm our proposal.

\begin{table}[htbp]
\centering
\caption{\label{tab:table1} The phonon PAM of fifteen phonon branches at the $K$ point of two In$_2$Se$_3$ materials with the opposite electric polarization $P$. For the $K'$ point, PAM has the opposite value.}
\resizebox{0.48\textwidth}{!}{
\begin{tabular}{c|rrrrrrrrrrrrrrrrr}
    \toprule
\diagbox [width=7.5em,trim=l] {polarization}{branch} & 1 & \makebox[0.02\textwidth][r]{2} & 3  & \makebox[0.02\textwidth][r]{4} & 5 & 6 & \makebox[0.02\textwidth][r]{7} & 8  & 9 & 10 & 11 & 12 & \makebox[0.02\textwidth][r]{13}  & \makebox[0.02\textwidth][r]{14} & 15\\
\hline
down ($-$P)& $-$1 & 0 & 1 & 0 & $-$1 & 1 & 0 & $-$1 & $-$1 & 1 & 1 & $-$1 & 0 & 0 & 1\\
up (+P)&  1 & 0 & $-$1 & 0 & 1 & $-$1 & 0 & 1 & 1 & $-$1 & $-$1 & 1 & 0 & 0 & $-$1\\
\toprule
\end{tabular}}
\end{table}

The In$_2$Se$_3$ crystal preserves $C_{3z}$ symmetry. In such a case, as mentioned, the chiral phonons at $C_{3z}$-invariant points, such as $K$ and $K'$ points of Brillouin zone, can have well defined PAM.
In Ref.~\cite{zhang2015chiral}, this PAM $\ell_{n\bm q}$ for the mode $\psi_{n\bm q}$ is defined as
\begin{equation}\label{PAM}
\mathcal{R}[(2\pi/3),z]|\psi_{n\bm q}\rangle= e^{-i(2\pi/3)\ell_{n\bm q}} |\psi_{n\bm q}\rangle,
\end{equation}
where $\mathcal{R}[(2\pi/3)]$ is the threefold rotation operator acting on the phonon mode, and $\bm q=K,K'$ for In$_2$Se$_3$ considered here. The computed PAM values for the 15 phonon modes at $K$ point are listed in Table.~\ref{tab:table1}. One observes that they indeed exhibit a sign change under ferroelectric switching. In previous studies, $\ell_{n\bm q}$ is often decomposed into an orbital (intercell) contribution and a spin (intracell) contribution. We find that both contributions change sign between the two polarization states. As symmetry eigenvalues, this PAM determines selection rules when chiral phonons interact with other quasiparticles, such as electrons and photons in optical transitions. Therefore, the electric control of phonon chirality also provides us a method to control other physical processes in ferroelectrics.

\begin{figure}[htbp]
\includegraphics[width=3.4 in,angle=0]{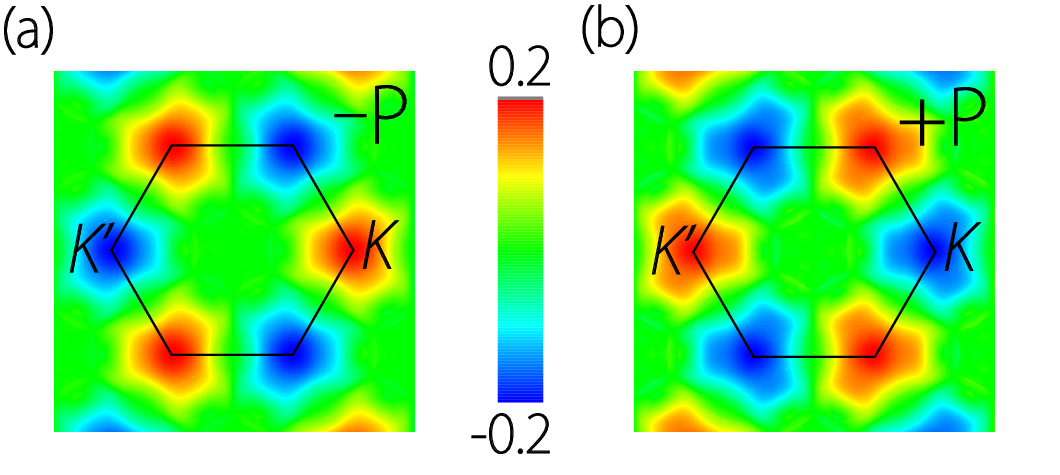}
\caption{\label{fig3} (a) and (b) show the $z$-direction phonon Berry curvature of the lowest (first) branch for the $-P$ and $+P$ In$_2$Se$_3$, respectively. The high-symmetry points $K$ and $K'$ are marked.}
\end{figure}

Another important quantity that is flipped during ferroelectric switching is phonon Berry curvature. For a 2D system, it is a pseudo-scalar (i.e., only with out-of-plane component). Similar to the Berry curvature for electronic band structure, it can be calculated by
\begin{equation}
  \Omega_{n\bm q}^z=-2\text{Im}\sum_{m\neq n}\frac{\langle \psi_{n\bm q}|\frac{\partial D}{\partial k_x}|\psi_{m\bm q}\rangle \langle \psi_{m\bm q}|\frac{\partial D}{\partial k_y}|\psi_{n\bm q}\rangle}{(\omega_{n\bm q}^2-\omega_{m\bm q}^2)^2},
\end{equation}
where $D(\bm q)$ is the dynamical matrix for the system, and $\omega_{n\bm q}$ is the eigen-frequency for the mode $\psi_{n\bm q}$. In Fig.~\ref{fig3}, we plot the calculated phonon Berry curvature distribution for the lowest ($n=1$) branch for the two polarization states. The result confirms that the Berry curvature for chiral phonons can also be electrically controlled via the ferroelectric switching.

\textcolor{blue}{\textit{Proposed Experimental Detection} ---}
Based on the above results, next, we propose effects that can be probed in experiment to demonstrate the electric control of chiral phonons.

\begin{figure}[tb]
\includegraphics[width=1\linewidth]{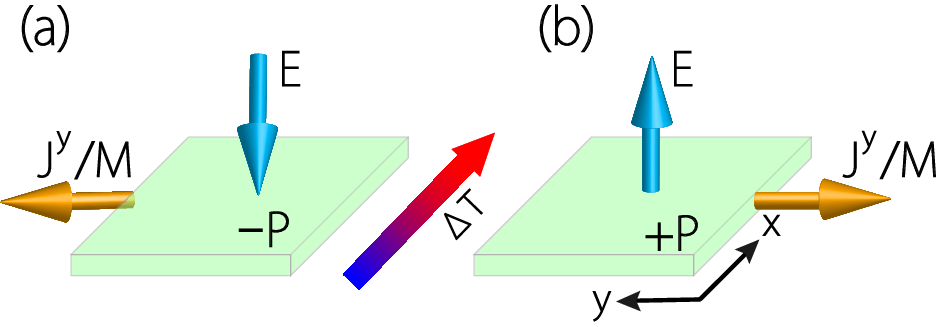}
\caption{\label{fig4} Schematic diagram of the phonon angular momentum/magnetization direction for $-P$ and $+P$ In$_2$Se$_3$ under the $x$-direction temperature gradient $\Delta T$, where $J^y$ represents the $y$-direction phonon angular momentum and $M$ represents the magnetization, $E$ represents the external electric field.}
\end{figure}

First, as we have mentioned, $S^i_{n\bm q}$ corresponds to the angular momentum of the phonon mode. Summing over all occupied modes, we would obtain the total angular momentum for phonons. In equilibrium, this is given by~\cite{zhang2014angular}
\begin{equation}\label{J_ph}	
J^i=\frac{1}{V}\sum_{n,\bm q}S^i_{n\bm q}\bigg[f_{0}(\omega_{n\bm q})+\frac{1}{2}\bigg],
\end{equation}
where  $f_{0}(\omega_{n\bm q})=1/(e^{\hbar\omega_{n\bm q}/k_{B}T}-1)$ is the Bose-Einstein distribution, $V$ is the volume of the sample, and $T$ is the temperature. Due to $\mathcal{T}$ symmetry, $S^i_{n\bm q}$ is opposite for modes at $\pm q$. Thus, the sum vanishes in equilibrium. On the other hand, $J^i$ could be nonzero when the chiral phonons are driven out of equilibrium. To achieve this, the most direct approach is to impose a temperature gradient $\nabla T$. Then, from Boltzmann equation, one easily finds the non-equilibrium distribution
\begin{equation}
  f_{n\bm q}=f_{0}-\tau\bm v_{n\bm q}\cdot \nabla T\frac{\partial f_{0}}{\partial T},
\end{equation}
where $\tau$ is the phonon relaxation time and $\bm v$ is the phonon group velocity. The induced phonon angular momentum is given by~\cite{hamada2018phonon}
\begin{equation}\label{J_temp}
J^i=-\frac{\tau}{V}\sum\limits_{n,\bm q}S^i_{n\bm q}\bm v_{n\bm q}\cdot \nabla T \frac{\partial f_{0}}{\partial T}=\beta^{ij}\partial_j T,
\end{equation}
where  $\beta^{ij}$ is a response tensor characterizing this thermally induced phonon angular momentum. For 2D In$_2$Se$_3$ with a $C_{3v}$ point group, we find that $\beta^{ij}$ must take the following simple form
\begin{equation}
  [\beta]=\left(
\begin{array}{ccc}
0& \beta^{xy} & 0 \\
-\beta^{xy}& 0 & 0 \\
0& 0 & 0
\end{array}
\right ).
\end{equation}
This means that for a temperature gradient along the $x$ direction, there will be a net phonon angular momentum induced along the $y$ direction. In the $-P$ polarization state, our calculation finds that 2D In$_2$Se$_3$ has $\beta^{xy}\sim 10^{-5}\times[\tau/(1\text{s})]\,\text{J\,s\,m}^{-2}\text{K}^{-1}$ at $T=300\, \text{K}$. Meanwhile, for the $+P$ polarization state, this $\beta$ tensor will flip sign. This induced phonon angular momentum is generally accompanied with a magnetic moment, since the oscillating ions are charged (in a simple picture described by the Born effective charge). For example, the gyromagnetic ratio $\gamma$ tensors of
In and Se ions can be expressed as $\gamma_{\alpha\beta}^{\text{In}}=g_\text{In}Z_{\alpha\beta}^\text{In}/2m_{\text{In}}$ and $\gamma_{\alpha\beta}^{\text{Se}}=g_\text{Se}Z_{\alpha\beta}^\text{Se}/2m_{\text{Se}}$, where the $g$'s are g-factors, the $Z$'s are Born effective charge tensors, and the $m$'s are the corresponding ion masses. Using the Born effective charge results from Ref.~\cite{soleimani2020ferroelectricity} and assuming g-factors on the scale of $1$-$10$, we obtain a rough estimate of the magnitude of the induced magnetization as
\begin{equation}
M\sim \frac{(\Delta T/L)}{(1\, \text{K}/\text{m})}\times 10^{-11} \text{A/m}.
\end{equation}
With current experimental technology, a temperature gradient of $10^{6}$-$10^{7}\ \text{K/m}$ can be achieved. This will lead to an induced magnetization on the order of $10^{-5}$-$10^{-4}$ A/m. Such a signal is large enough to be detected with Magneto-optical Kerr technique. Here, in 2D In$_2$Se$_3$, the most important feature is that under ferromagnetic switching driven by the $E$ field, this thermally induced magnetization must also get reversed, as illustrated in Fig.~\ref{fig4}. The detection of this reversal can serve as an experimental proof of our idea.

%\subsection{Tunable Phonon Berry curvture and Hall effect}

\begin{figure}[tb]
\includegraphics[width=3.2 in,angle=0]{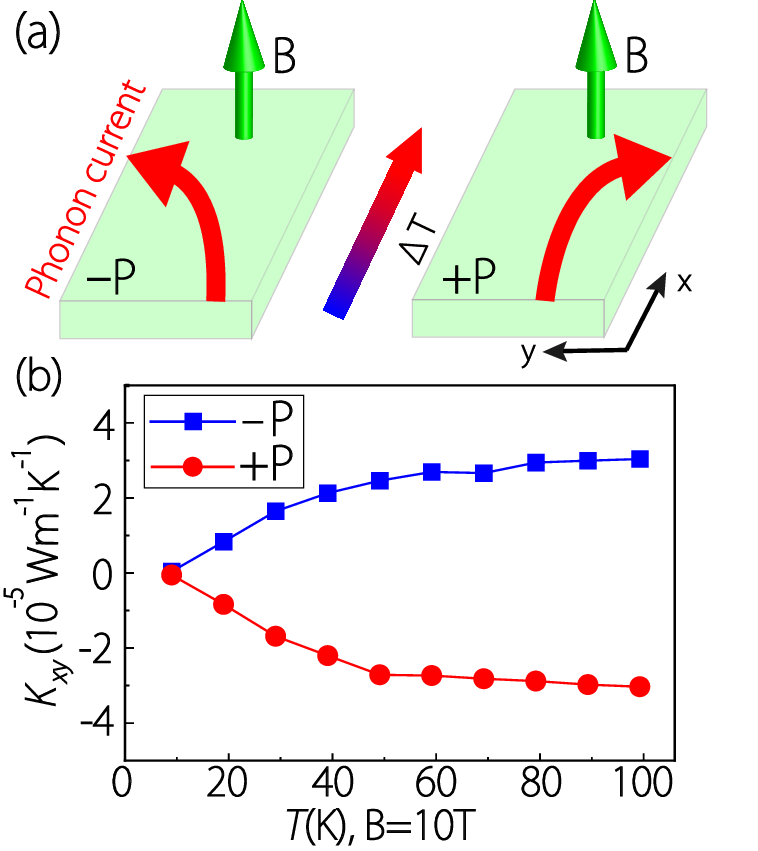}
\caption{\label{fig5} (a) Schematic diagram of phonon Hall effect under the temperature gradient and magnetic field, where $B$ represents the applied magnetic field, and $-P$ and $+P$ represent the polarization direction of the ferroelectric material In$_2$Se$_3$. The red arrows indicate the deflection of Hall phonon current. (b) Calculated phonon Hall conductivity versus temperature at $B = 10$ T.}
\end{figure}

The second effect we propose is an electrically tunable phonon Hall effect. The phonon Hall effect has been studied in several systems in the past. In linear response, the effect requires the broken of $\mathcal{T}$ symmetry. In previous experiments, this was achieved by an applied strong $B$ field. It was shown that the phonon Berry curvature can give rise to an intrinsic contribution to the Hall transport. The corresponding phonon Hall thermal conductivity can be expressed as~\cite{qin2012berry,PhysRevB.103.214301}

\begin{equation}\label{kappa}
\kappa_{xy}=\frac{1}{2\hbar T}\int^{\infty}_{-\infty} d\epsilon\ \epsilon^{2}\eta(\epsilon)\frac{\partial f_0(\epsilon)}{\partial\epsilon},
\end{equation}
where
\begin{equation}
    \eta(\epsilon)=\frac{1}{V}\sum_{n,\bm q}\Omega_{n\bm{q}}^{z}\Theta(\epsilon-\hbar\omega_{n\bm q}),
\end{equation}
and $\Theta$ is the step function. In Fig.~\ref{fig5}, we plot the calculated $\kappa_{xy}$ as a function of temperature for the two polarization states of In$_2$Se$_3$ under an applied $B$ field of 10 T. Here, the influence of $B$ field on phonons is through spin-phonon coupling by using the approach in Ref~\cite{PhysRevB.103.214301}. One observes that the two states indeed have $\kappa_{xy}$ with opposite signs. In other words, the phonon Hall current will reverse its direction under ferroelectric switching. One also note that due to $\mathcal{T}$ symmetry breaking, the magnitudes of $\kappa_{xy}$ for $+P$ and $-P$ states are not exactly the same. However, the difference is small, because of the generally weak influence of $B$ field on phonon system, as we mentioned at the beginning. Experimentally, $\kappa_{xy}$ with values down to $10^{-6}\ \text{Wm}^{-1}\text{K}^{-1}$ has been detected~\cite{strohm2005phenomenological}. Therefore, we expect our predicted effect in 2D ferroelectric In$_2$Se$_3$ should also be detectable.

%\section{discussion and conclusion}

\textcolor{blue}{\textit{Discussion and Conclusion} ---}
Finally, we mention the possibility to realize topological interface phonon modes at ferroelectric domain boundaries. Consider a 2D ferroelectric with hexagonal lattice structure. If the phonon bands at $K$ point form Dirac type valley structure, the integral of Berry curvature of such valley bands will give a valley charge $N_v$, similar to that in gapped graphene. Now, because the two polarization states have opposite Berry curvatures, they must also have opposite $N_v$. For a  boundary between two ferroelectric domains (where the two valleys can be resolved), the change of $N_v$ across the boundary, $\Delta N=2N_v$, gives the number of topological interface chiral phonon mode at $K$ valley. Meanwhile, due to $\mathcal{T}$, there are also $\Delta N$ chiral modes at $K'$ valley but they propagate in the opposite direction. While the above physical mechanism is general, the advantage for the current system is that the ferroelectric domains and their boundaries can be well controlled by applied $E$ field, e.g., through gating technique. In this way, one can in principle manipulate and steer 1D phonon propagation in real time. Having said these, we note that to find a material platform with such Dirac valley phonon bands is not trivial. For example, 2D In$_2$Se$_3$ does not have such a feature.
It remains an interesting task to search for suitable ferroelectric materials to realize such interface phonon modes in future studies.

In conclusion, we have proposed a new route to control chiral phonons by using ferroelectric material platforms. We show that applied electric field can flip phonon chirality, PAM, and Berry curvature through ferroelectric switching. The switch of these key characters for chiral phonons can lead to a range of interesting effects with electric controllability, such as thermally induced magnetization, phonon Hall effect, and interface phonon modes at ferroelectric domain boundaries. We have demonstrated our ideas in a concrete material platform, the 2D In$_2$Se$_3$, which yields experimentally measurable results. These findings will be useful for future development of thermal and information devices based on chiral phonons.

\textcolor{blue}{\textit{Acknowledgement} ---}
This work was funded by NSFC (NO.12247149), China Postdoctoral Science Foundation 2023M733410, and Singapore Ministry of Education AcRF Tier 2 (T2EP50220-0026). Hao Chen thanks Qianqian Wang, Xunkun Feng, Ning Ding, Wen He and Chen Shen for the helpful discussions.

\bibliography{aps-main}

\end{document}